\documentstyle{article}

\input{tcilatex}
\begin{document}

\title{{\bf BARYON MASS SPECTRA WITH QCD - TYPE SCREENED POTENTIAL}}
\author{{\bf V. Rubish} \\
Department of Theoretical Physics, Uzhgorod State University, 32 Voloshyn
Str., UA-88000}

\begin{titlepage}
\vskip 0.3cm

\centerline{\large \bf Baryon mass spectra}
\centerline{\large \bf with QCD- type screened potential}

\vskip 0.7cm

\centerline{V. Rubish}

\vskip .3cm

\centerline{\sl Uzhgorod State University,}
\centerline{\sl Department of Theoretical Physics, Voloshin str. 32,}
\centerline{\sl 88000 Uzhgorod, Ukraine}

\vskip 1.5cm

\begin{abstract}
The work is devoted to the description of baryon mass spectrum in the 
quasy-relativistic quark potential model. For cosidering 3 - quark system
 advantageous is the use of hyperspherical functions. For obtaining the
 mass - seguence of $\Delta _{33}-$isobar the numerical computations
 were carryed out with the use of screened QCD - motivated potential, 
the use of which gives quite acceptable results even in the nonrelativistic
 approximation.
\end{abstract}

\vskip .3cm

\vskip 8cm

\hrule

\vskip .9cm

\noindent
\vfill $ \begin{array}{ll} ^{\ast}\mbox{{\it e-mail
address:}} &
 \mbox{vrubish@univ.uzhgorod.ua}
\end{array}
$

\vfill
\end{titlepage}\eject
\baselineskip=14pt

It is widely accepted that quark potential model is able to give a rather
good description of heavy quark systems, though, the problem of application
of such a model to the light quark systems,especially to baryon case is more
complicated. On the other hand, the study of such systems is attractive for
many reasons. Firstly, the experimental situation in this case is much more
abundant. Secondly, there is no hope in near future to solve the
relativistic three-body problem. Moreover, the baryon spectra are usualy
treated separately from the meson ones and so the question arises how far
the barions and mesons (both light and heavy ones) can be treated in the
unified approach.

Usually the potential which is used nas the power-law form

\begin{equation}
V\left( r\right) =-\frac{\alpha _{s}}{r}+Ar^{\gamma },\text{ }\gamma =1,2,2/3
\label{f1}
\end{equation}

Of these potentials the most appropriate seems to be potential with $\gamma $
= 1 called Cornell - potential. The attempt to compare all these potentials
is given in \cite{1} where the necessary references to previous works is
also given.

On the other hand involving the virtual loop - diagram of fermions in
lattice gauge calculation leads to colour screening effects - the potential
does not rise at very large separation. Clearly, these colour screaning
effects may reveal themselves in the highly - excited barion resonances.
Meantime, a very promising QCD - motivated potential was proposed by
Chikovani, Jenkovszky and Paccanoni (CJP) on the basis of the model of a non
- perturbative gluon propagator. It was succesfully applied to the heavy
meson spectroscopy \cite{2} and it turns out that such a potential having
the form

\begin{equation}
V\left( r\right) =\frac{g^{2}}{6\pi \mu }\left( 1-e^{-\mu \cdot r}\right) -%
\frac{16\pi }{25}\frac{e^{-k\cdot r}}{\ln \left( b+\frac{1}{\left( 
\widetilde{\Lambda \cdot }r\right) ^{2}}\right) }  \label{f2}
\end{equation}
contains color screening effects. None of the above mentioned power - law
potentials posses the screening property. Previously, the colour screaning
effects in baryon spectroscopy were studied in \cite{3} using similar, but
purely phenomenological potentials. In the present paper we shall study the
baryon mass - spectra using the CJP potential.

To obtain the mass - spectrum of the three - quark problem one has to solve
the eigenvalue problem, which in our model means solving the Schr\"{o}dinger
equation with the appropriate boundary conditions. To determine the wave
function of relative motion $\varphi \left( \rho ,\tau \right) $ we go over
to the spherical coordinate system in the six - dimensional configuration
spece, which is often referred to as hyperspherical coordinates.

As was disscused in \cite{1} it turns out sufficient to restrict ourselves
to K - zero approximation in which case the Schr\"{o}dinger equation is
reduced to a single radial equation:

\begin{equation}
\left[ \frac{d^{2}}{dr^{2}}-\frac{K\left( K+1\right) }{r^{2}}-2\mu \left(
E-\left\langle \Phi _{0}\left| V\left( R,\alpha \right) \right| \Phi
_{0}\right\rangle \right) \right] u\left( R\right) =0.  \label{f3}
\end{equation}

Here 
\begin{equation}
u\left( R\right) =R^{-3/2}\cdot \chi \left( R\right) ,\text{ }K=\Lambda +3/2.
\label{f4}
\end{equation}
The matrix elements $\left\langle \Phi _{0}\left| V\left( R,\alpha \right)
\right| \Phi _{0}\right\rangle $, according to \cite{1}, assume the form

\begin{equation}
W_{n}=\frac{16}{\pi }\int\nolimits_{0}^{\pi /2}V\left( r_{ij}\right) \cdot
\sin ^{2}\alpha \cdot \cos ^{2}\alpha \cdot d\alpha .  \label{f5}
\end{equation}

The mass of three - quark system is calculated according to the relation

\begin{equation}
M=3m_{q}+E+V_{0},  \label{f6}
\end{equation}

Here $V_{0}$ is additional parameter. Such parameters are present in QCD -
motivated potentials and describe the energy level shift in QCD - vacuum.
Besides $V_{0}$ can include in our model spin - dependent and relativistic
corrections (like $p^{2}/8m^{3}$) and, thus, it can depend on quark mass.
The discussion concerning the meaning of $V_{0}$ is given also in \cite{4}.

In what follows we have used the following values of the parameters. The
masses of up and down quarks were taken to be as usually in quark models 
\cite{1}, \cite{4} $m_{u}=m_{d}=0.33$ $GeV$ wich follows from magnetic
momentum considerations. For CJP potential the parameters were choosen as: $%
g^{2}/6\pi =0.24$ $GeV^{2}$, $\mu =0.054$ $GeV^{2}$, $\Lambda =0.35$ $GeV,$ $%
k=0.75$ $GeV^{2}$, $b=4$ \cite{2}. While $V_{0}$ is chosen to be equal $%
V_{0}=1.376$ $GeV$.

In the present paper, we have considered $\Delta _{33}$-trajectories with
the following quantum numbers $P=+1,$ $J=3$/2, 7/2, 11/2,... and with
signature $\sigma =1$. The results of calculations together with
experimental data are presented in table 1. The experimental data are taken
from \cite{5}.

\begin{tabular}{|c|c|c|c|c|c|}
\hline
$N$ & $L$ & $\left( 3/2\right) ^{+}$ & $\left( 7/2\right) ^{+}$ & $\left(
11/2\right) ^{+}$ & $\left( 15/2\right) ^{+}$ \\ \hline
$0$ & $M_{EXP}$ & $1232\pm 2$ & $1950\pm 10$ & $2420\pm 80$ & $2950\pm 70$
\\ \hline
& $CJP$ & $1223$ & $1953$ & $2479$ & $2949$ \\ \hline
$1$ & $M_{EXP}$ & $1920\pm 80$ & $2390\pm 38$ &  &  \\ \hline
& $CJP$ & $1813$ & $2397$ &  & $\chi ^{2}=1.35$ \\ \hline
$2$ & $M_{EXP}$ & $-$ &  &  &  \\ \hline
& $CJP$ & $2290$ &  &  &  \\ \hline
\end{tabular}

Table 1. Baryon count, $\Delta _{33}$-trajectory (MeV)

As one can see from the table 1, the description of the existing data given
by CJP potential is very good. So we can see that in addition to $q\overline{%
q}-$systems \cite{6} the screened QCD-type potential gives also the
description of $qqq-$systems, i.e. baryons.

\begin{center}
\_\_\_\_\_\_\_\_\_\_\_\_\_\_\_\_\_\_\_\_\_\_\_\_\_\_\_\_\_\_\_\_\_\_\_\_\_\_%
\_\_\_\_\_\_\_\_\_
\end{center}


\begin{thebibliography}{9}
\bibitem{1}  V. Lengyel, M. Haysak, Ukr. Phys. Journ. {\bf vol.37 }N9 (1992)
1287.

\bibitem{2}  Z.E. Chikovani, L.L. Jenkovszky, F.Paccanoni, Mod. Phys. Lett. 
{\bf B 173} (1986) 437.

\bibitem{3}  Dong Yubing, Yu You-wen, Preprint BIHEP-TH-92-27, Beijing,
China, 1992; see also Zhand Zong-ye, Yu You-wen, Shen Peng-nian, Preprint
BIHEP-TH-92-28 Beijing, China, 1992.

\bibitem{4}  Wolfgang Lucha, Franz F. Sch\"{o}berl:  {\it Die Starke
Wechselwirkung. Eine Einf\"{u}hrung in nichtrelativistische Potential
modelle,} \copyright Bibliographisches Institut \& F.A.Brockhaus AG,
Mannheim 1989.

\bibitem{5}  Part. Data Group, Phys. Rev. {\bf D 1 }(1996).

\bibitem{6}  V. Lengyel, V. Rubish, Yu. Fekete, S.Chalupka, M. Salak,
Journ.of Phys. Stud.,1, {\bf vol.2} (1998) 38.
\end{thebibliography}
\end{document}